\documentclass[superscriptaddress,twocolumn,showpacs,prb,floatfix]{revtex4}
\usepackage{epsfig}
\usepackage{tabularx}

\newcommand{\be}{\begin{equation}}
\newcommand{\ee}{\end{equation}}
\newcommand{\beqn}{\begin{eqnarray}}
\newcommand{\eeqn}{\end{eqnarray}}
\newcommand{\nsw}{N_{\mathrm{sweep}}}
\newcommand{\nsa}{N_{\mathrm{samp}}}

\begin{document}

\title{Geometry of large-scale low-energy excitations in 
the one-dimensional Ising spin glass with power-law interactions}

\author{Helmut G.~Katzgraber}
\affiliation{Theoretische Physik, ETH H\"onggerberg, 
CH-8093 Z\"urich, Switzerland}

\author{A.~P.~Young}
\email{peter@bartok.ucsc.edu}
\homepage{http://bartok.ucsc.edu/peter}
\affiliation{Department of Physics, University of California,
Santa Cruz, California 95064, USA}

\date{\today}

\begin{abstract}
Results are presented for the geometry of low-energy excitations in the
one-dimensional Ising spin chain with power-law interactions, in which the
model parameters are chosen to yield a finite spin-glass transition
temperature. Both finite-temperature and ground-state studies are carried
out.  For the range of sizes studied the data cannot
be fitted to any of the standard spin-glass scenarios without including
corrections to scaling. Incorporating such corrections we find that the
fractal dimension of the surface of the excitations, is either
equal to the space dimension, consistent with replica symmetry
breaking, or very slightly less than it.  The latter case is consistent
with the droplet and ``trivial-nontrivial'' pictures.
\end{abstract}

\maketitle

\section{Introduction}
\label{introduction}

There have been several numerical attempts at
finite temperature\cite{krzakala:00,palassini:00,marinari:00,katzgraber:01,katzgraber:01a,katzgraber:01c,houdayer:00,houdayer:00b}
and
zero temperature\cite{bray:84,mcmillan:84,mcmillan:84b,rieger:96,hartmann:99,palassini:99,hartmann:01a,carter:02}
to better understand the nature of the spin-glass state for short-range
spin glasses. These results are generally interpreted in terms of the two
main theories for the spin-glass phase: the replica symmetry-breaking
(RSB)  picture,\cite{parisi:79,parisi:80,parisi:83,mezard:87} and the
``droplet picture''\cite{fisher:86,fisher:87,fisher:88,bray:86} (DP). RSB
predicts that excitations involving a finite fraction of the spins cost
only a finite energy in the thermodynamic limit, and that the fractal
dimension of the \textit{surface} of these excitations $d_s$ is equal
to the space dimension $d$. This is in contrast to DP where a low-energy
excitation (droplet) has an energy proportional to $\ell^{\theta}$, where
$\ell$ is the characteristic length scale of the droplet and $\theta$ is a
positive stiffness exponent. In addition, the surface of these excitations
is fractal with $d_s < d$.  More recently Krzakala and
Martin,\cite{krzakala:00} as well as Palassini and
Young,\cite{palassini:00} suggest an intermediate picture (called ``TNT''
for trivial-nontrivial) in which droplets have a fractal surface with $d_s
< d$, and their energy is finite in the thermodynamic limit. Which of the
above pictures describes the spin-glass state correctly is still widely
debated.

The RSB and TNT pictures require two stiffness exponents for the energy of
large-scale excitations. There is convincing numerical evidence that
changing the boundary conditions (e.g., from periodic to antiperiodic),
which induces a \textit{domain wall}, costs an energy which increases as
$\ell^\theta$ with $\theta > 0$. On the other hand, in the RSB and TNT
pictures, the energy of \textit{droplets}, created by thermal noise or by
applying a perturbation for a fixed set of boundary conditions, varies as
$\ell^{\theta'}$ with $\theta' = 0$. By contrast, the DP makes the
reasonable ansatz that $\theta' = \theta\ (>0)$.

In a previous publication,\cite{katzgraber:03} we studied the
one-dimensional long-range Ising spin glass with power-law interactions.
The model's advantage is that large system sizes can be studied, in
contrast to the short-range spin-glass models commonly used. The results
of Ref.~\onlinecite{katzgraber:03} showed that the stiffness exponent
$\theta$ for zero-temperature domain-wall excitations is positive and in
fair agreement with analytical predictions from the droplet
model.\cite{fisher:88,bray:86b,katzgraber:03} However, the stiffness
exponent $\theta^\prime$ for thermally induced droplet excitations is
different and consistent with zero.  Hence, \textit{at least for the range
of system sizes studied}, $L \le 512$, the data of
Ref.~\onlinecite{katzgraber:03} are consistent with both the TNT and RSB
scenarios since they have $\theta > \theta^\prime = 0$.

The purpose of the present paper is to estimate $d_s$, because this
distinguishes between the RSB and TNT scenarios, since $d_s = d$ in RSB
while $d_s < d$ in TNT. For short-range models, a droplet excitation forms
a single connected piece, and so $d_s$ has to be zero in $d=1$. However, for
long-range interactions, a droplet may consist of disconnected
pieces,\cite{fisher:88} so a nontrivial value of $d_s$ is possible in $d
= 1$.  We perform both finite-temperature Monte Carlo simulations and
ground-state studies. Our results suggest that droplets are possibly
compact in agreement with RSB, although the data are also consistent with a
very small value of $d - d_s$, which would be consistent with TNT.

In Sec.~\ref{model} we introduce the model, observables, and details of
the Monte Carlo technique. Results at zero temperature are presented in
Sec.~\ref{Teq0res}, and those at finite temperature are presented in
Sec.~\ref{Tgt0res}. Our conclusions are summarized in
Sec.~\ref{conclusions}.

\section{Model and Numerical Method}
\label{model}

The Hamiltonian of the one-dimensional long-range Ising spin chain with
power-law interactions is given by
\begin{equation}
{\cal H} = -\sum_{\langle i,j\rangle} J_{ij} S_i S_j \; ,
\label{hamiltonian}
\end{equation}
where the Ising spins $S_i = \pm 1$ are evenly distributed on a circular
ring of length $L$ to ensure periodic boundary conditions. The sum is over
all pairs of spins on the chain and the couplings $J_{ij}$ are given by
\begin{equation}
J_{ij} = c(\sigma)\frac{\epsilon_{ij}}{r_{ij}^\sigma}\; ,
\label{bonds}
\end{equation}
where\cite{katzgraber:03}
\begin{equation}
r_{ij} = \frac{L}{\pi} \sin\left({\pi |i-j| \over L}\right)
\end{equation}
is the straight-line distance between sites $i$ and $j$. The random part
of the interactions $\epsilon_{ij}$ is chosen according to a Gaussian
distribution with zero mean and standard deviation unity, and the constant
$c(\sigma)$ in Eq.~(\ref{bonds}) is chosen\cite{katzgraber:03} to give a
mean-field transition temperature $T_c^{\rm MF} = 1$.

The one-dimensional long-range Ising spin chain has a very rich phase
diagram\cite{fisher:88,bray:86b,katzgraber:03} in the $d$-$\sigma$ plane.
Spin-glass behavior is controlled by the long-range part of the
interaction if $\sigma$ is sufficiently small, and by the short-range part
if $\sigma$ is sufficiently large. In this work we focus on the long-range
behavior at $\sigma = 0.75$ for which\cite{katzgraber:03} $T_c > 0$ and
the critical exponents are non-mean-field like. Using the exact
relation\cite{bray:86b,fisher:88} $\theta = d - \sigma$, we expect $\theta
= 0.25$ for $d = 1$, which is in moderate agreement with numerical
results\cite{katzgraber:03} for \textit{domain walls} induced by a change
in boundary conditions at $T=0$. By studying thermally induced
\textit{droplet} excitations Ref.~\onlinecite{katzgraber:03} also
estimated $\theta^\prime \approx 0$, consistent with RSB and TNT.

In order to excite droplets at zero temperature we use the
coupling-dependent ground-state perturbation method described
elsewhere.\cite{palassini:00,palassini:02,hartmann:02} First, we compute
the ground-state configuration $\{S_i^{(0)}\}$. Then we perturb the
couplings $J_{ij}$ by the following amount:
\begin{equation}
\Delta{\mathcal H}(\epsilon) = {2 \epsilon \over N}\sum_{\langle i,j\rangle}
{[J_{ij}^2]_{\rm av} \over (T_c^{MF})^2} 
S_i^{(0)} S_j  ^{(0)} S_i S_j \; ,
\label{delta_h}
\end{equation}
where $\epsilon$ is a coupling constant and $[\cdots]_{\rm av}$ represents
a disorder average. The (total) energy of the unperturbed ground state
then increases by exactly $\epsilon$, whereas the energy of any other
state $\alpha$ will increase by the lesser amount of $\epsilon q_l$, where
$q_l$ is the link overlap between the unperturbed ground state and a state
$\alpha$:
\begin{equation}
q_l = {2 \over N}\sum_{\langle i,j\rangle} {[J_{ij}^2]_{\rm av} \over
(T_c^{MF})^2} 
S^{(\alpha)}_i S^{(\alpha)}_j 
S^{(0)}_i S^{(0)}_j  \; .
\label{ql0}
\end{equation}
In previous work $q_l$ has been defined for nearest-neighbor models in
which the sum is over nearest-neighbor pairs. Here we have generalized the
link overlap to long-range models in a natural way.  Because the coupling
constant $\epsilon$ is of order unity and not of order $L$ only 
{\em low-energy} excitations can be generated. We compute the new ground state 
of the perturbed system and record the link overlap between the old and new
ground states. In the context of zero-temperature simulations the term
``link overlap'' will hereafter refer to the link overlap between the
perturbed and unperturbed ground states.

Ground states are calculated using the parallel tempering Monte Carlo
method\cite{hukushima:96,marinari:98b} (at very low temperatures) as
described in Refs.~\onlinecite{katzgraber:03} and \onlinecite{moreno:02}.
The parameters used in the $T = 0$ simulations are shown in Table
\ref{simparamst0}. For each value of $L$ and $\epsilon$ we compute $10^4$
disorder realizations. We find that for $\sigma = 0.75$, when the model is
in the long-range phase, the efficiency of the used algorithm to calculate
ground states scales as $L^z$ with $z = 2.9 \pm 0.3$. For the current
project this translates to a total CPU time of 70 CPU years. Curiously,
for $\sigma = 2.50$, for which the interactions are effectively short
range so frustration is minimal in the $d=1$ model studied here, the
algorithm performs poorly with the equilibration time varying as $\sim
\exp(aL)$, with $a = 0.13 \pm 0.02$. It would be useful to understand
intuitively the reasons for this.

\begin{table}
\caption{
Parameters of the $T = 0$ simulations. The table shows the total number of
Monte Carlo steps used for each value of $\epsilon$ and $L$. All data are
computed with $10^4$ disorder realizations. The lowest temperature used to
calculate the ground states with parallel tempering Monte Carlo is $T = 0.05$, 
the highest $1.70$. We use between 10 and 23 temperatures, depending on 
the system size, to ensure that the acceptance ratios of the parallel 
tempering moves are larger than $\sim 0.30$.
\label{simparamst0}
}
\begin{tabular*}{\columnwidth}{@{\extracolsep{\fill}} c c c c c c c}
\hline
\hline
$\epsilon$ &  $L = 16$  & $L = 32$ & $L = 64$ & $L = 128$ & $L = 256$ &
$L = 512$\\
\hline
0.50 & $2\times 10^3$ & $4\times 10^3$ & $8\times 10^3$ & $4.0\times 10^4$ &
$3.2\times 10^5$ & $4.0 \times 10^5$\\
1.00 & $2\times 10^3$ & $4\times 10^3$ & $8\times 10^3$ & $4.0\times 10^4$ &
$3.2\times 10^5$ & $4.0 \times 10^5$\\
2.00 & $2\times 10^3$ & $4\times 10^3$ & $8\times 10^3$ & $4.0\times 10^4$ &
$3.2\times 10^5$ & $4.0 \times 10^5$\\
\hline
\hline
\end{tabular*}
\end{table}

One quantity that we study at $T = 0$ is the link overlap, averaged over
\textit{all samples}. To see how this varies with size\cite{palassini:00}
consider a large cluster of excited spins. This has a characteristic energy
of order $\sim L^{\theta^\prime}$, which is to be compared with the energy
gained from the perturbation $\epsilon (1 - q_l) \sim \epsilon L^{-(d -
d_s)} $. There is a distribution of cluster energies which we assume to
have a finite weight at the origin, so the probability that the
perturbation will create the excitation is $\sim \epsilon
L^{-(\theta^\prime + d - d_s)}$. When this occurs $1 - q_l \sim L^{-(d -
d_s)} $, and so on average\cite{palassini:00,palassini:02,hartmann:02}
\begin{equation}
[1 - q_l]_{\rm av} = \epsilon L^{-\mu_l}(a + bL^{-c})\; ,
\label{ql_scale}
\end{equation} 
where
\begin{equation}
\mu_l = \theta^\prime + 2(d - d_s) 
\label{eq_mul}
\end{equation} 
and we have added a correction to scaling term $bL^{-c}$.

In RSB we have $\mu_l = 0$ so Eq.~(\ref{ql_scale}) tends to a constant for
$L \to \infty$, whereas in DP and TNT $\mu_l > 0$ so Eq.~(\ref{ql_scale})
tends to zero in this limit.

In addition, we consider averages over \textit{only those samples in which
a large excitation is generated},\cite{hartmann:02,palassini:02}
comprising a finite fraction of spins. The criterion we take is $|q| \le
0.50$. Averaging just over these samples
gives\cite{hartmann:02,palassini:02}
\begin{equation}
[1 - q_l]^\prime_{\rm av} = L^{-(d - d_s)}(a + bL^{-c})\; ,
\label{qlp_scale}
\end{equation}
the prime representing the restricted average.  Equation (\ref{qlp_scale})
follows trivially from the arguments presented in the derivation of
Eq.~(\ref{ql_scale}) with the probability factor $\epsilon
L^{-(\theta^\prime + d - d_s)}$ replaced by unity. We expect that $[1 -
q_l]^\prime_{\rm av}$ will be independent of $\epsilon$.

In order to study droplet geometries at finite temperatures, we compute
the distribution of the link overlap $q_l$ between two replicas $\alpha$
and $\beta$ of the system with the same disorder:
\begin{equation}
q_l = {2 \over N}\sum_{\langle i,j\rangle} {[J_{ij}^2]_{\rm av} \over
(T_c^{MF})^2} 
[ \langle S^{(\alpha)}_i S^{(\alpha)}_j 
S^{(\beta)}_i S^{(\beta)}_j\rangle ]_{\rm av} \; .
\label{ql}
\end{equation}
Here $\langle \cdots \rangle$ represents a thermal average, and
$[\cdots]_{\rm av}$ represents a disorder average.  From the finite-size
scaling arguments\cite{katzgraber:01} used to derive Eq.~(\ref{ql_scale})
we expect that the variance of the distribution of the link overlap scales 
as
\begin{equation} 
{\rm Var}(q_l) = L^{-\mu_l} \left( a + bL^{-c} \right) \; .
\label{varql_scaling}
\end{equation}

Note that in RSB $\mu_l = 0$ so ${\rm Var}(q_l)$ tends to a constant for
$L \to \infty$. However, $\mu_l > 0$ in DP (since $\theta' = \theta > 0$
and $d_s < d$) and in TNT (since $d_s < d$). The $bL^{-c}$ term is a
correction to scaling, which turns out to be necessary since the data
cannot be fitted without it.

To speed up equilibration of the finite-$T$ simulations we use the
parallel tempering Monte Carlo method.\cite{hukushima:96,marinari:98b} We
test for equilibration using the criterion developed
earlier,\cite{katzgraber:01} now generalized\cite{katzgraber:03} for the
Hamiltonian in Eq.~(\ref{hamiltonian}). For all sizes, the lowest
temperature used is $T = 0.05$, well below $T_c \simeq
0.63$.\cite{katzgraber:03,leuzzi:99} The highest temperature is $1.70$
which is well above the mean-field critical temperature ( $T_c^{\rm
MF}=1$) and so the spins equilibrate fast there. We choose the spacing
between the temperatures such that the acceptance ratios for the global
moves are around $0.30$. Parameters of the finite-$T$ simulations are
summarized in Table~\ref{simparams}.

\begin{table}
\caption{
Parameters of the finite-$T$ simulations. $\nsa$ is the number of samples,
$\nsw$ is the total number of Monte Carlo sweeps for each of the $2 N_T$
replicas for a single sample, and $N_T$ is the number of temperatures used
in the parallel tempering method.
\label{simparams}
}
\begin{tabular*}{\columnwidth}{@{\extracolsep{\fill}} c r r l }
\hline
\hline
$L$  &  $\nsa$  & $\nsw$ & $N_T$  \\
\hline
16  & $2.0 \times 10^4$ & $2.0 \times 10^3$ & 10 \\
32  & $2.0 \times 10^4$ & $4.0 \times 10^3$ & 10 \\
64  & $2.0 \times 10^4$ & $8.0 \times 10^3$ & 12 \\
128 & $2.0 \times 10^4$ & $4.0 \times 10^4$ & 14 \\
256 & $1.0 \times 10^4$ & $2.0 \times 10^5$ & 17 \\
512 & $5.0 \times 10^3$ & $8.0 \times 10^5$ & 24 \\
\hline
\hline
\end{tabular*}
\end{table}

To summarize, for $L \to \infty$ all the quantities that we calculate, $[1
- q_l]_{\rm av}$ in Eq.~(\ref{ql_scale}), $[1 - q_l]^\prime_{\rm av}$ in
Eq.~(\ref{qlp_scale}), and ${\rm Var}(q_l)$ in Eq.~(\ref{varql_scaling})
tend to a non-zero constant in RSB, whereas they tend to zero with a power
of $L$ in TNT and DP.

\section{Results at Zero Temperature}
\label{Teq0res}

\begin{figure}
\centerline{\epsfxsize=\columnwidth \epsfbox{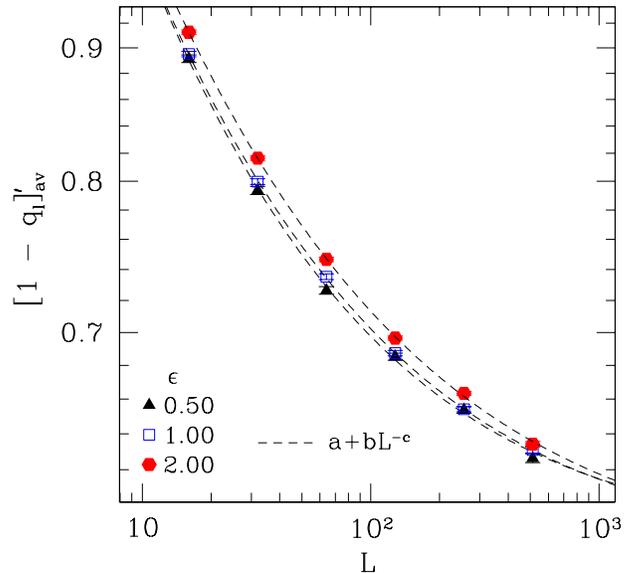}}
\vspace{-1.0cm}
\caption{
Zero-temperature data for $[1 - q_l]^{\prime}_{\rm av}$ as a function of
system size $L$ for different values of the coupling constant $\epsilon$.
Note that the data only depend slightly on $\epsilon$, thus indicating
only small deviations from the scaling form. The dashed lines correspond
to a three-parameter fit to $a + bL^{-c}$ as expected in RSB.
}
\label{omqlp_fig}
\end{figure}

We first discuss the results for the constrained average of $1 - q_l$,
including only samples $|q| \le 0.5$, since this yields $d - d_s$
independent of $\theta^\prime$, see Eq.~(\ref{qlp_scale}). The results are
shown in Fig.~\ref{omqlp_fig}. Note that the data only depend slightly on
$\epsilon$, indicating only small deviations from the expected scaling
form.

The results of a DP/TNT fit to $[1 - q_l]^{\prime}_{\rm av} = L^{-(d -
d_s)}(a + bL^{-c})$ are presented in Table \ref{table_omqlpfit1}, while
the corresponding RSB fits to $[1 - q_l]^{\prime}_{\rm av} = a + bL^{-c}$
are shown in Table \ref{table_omqlpfit2}. In the DP/TNT fits we find that
$d - d_s$ is close to zero.

\begin{table}
\caption{
Fits of zero-temperature data for $[1 -q_l]^{\prime}_{\rm av} $ to $
L^{-(d - d_s)}(a + bL^{-c})$, appropriate for DP/TNT, for different
coupling constants $\epsilon$. The last column is $\chi^2$ per degree of
freedom, where for this data with six points and four fitting parameters, the
number of degrees of freedom (ndf) is two.
\label{table_omqlpfit1}
}
\begin{tabular*}{\columnwidth}{@{\extracolsep{\fill}} l l l l l l}
\hline
\hline
$\epsilon$ & $d - d_s$ & $a$ & $b$ & $c$ & $\chi^2/\mbox{ndf}$ \\
\hline
$0.50$ & $0.043(14)$ & $0.81(8)$  & $1.95(64)$ & $0.83(19)$ & $0.64$\\
$1.00$ & $0.003(28)$ & $0.59(14)$ & $1.29(7)$  & $0.51(9)$  & $0.31$\\
$2.00$ & $0.019(19)$ & $0.67(10)$ & $1.30(8)$  & $0.54(8)$ & $1.63$\\
\hline
\hline
\end{tabular*}
\end{table}

\begin{table}
\caption{
Fits of zero-temperature data for $[1 - q_l]^{\prime}_{\rm av} = a +
bL^{-c}$, appropriate for RSB, for different values of the coupling
constant $\epsilon$. The number of degrees of freedom (ndf) here is three.
\label{table_omqlpfit2}
}
\begin{tabular*}{\columnwidth}{@{\extracolsep{\fill}} l l l l l}
\hline
\hline
$\epsilon$ & $a$ & $b$ & $c$ & $\chi^2/\mbox{ndf}$\\
\hline
$0.50$ & $0.580(7)$ & $1.35(8)$ & $0.53(3)$ & $1.74$ \\
$1.00$ & $0.575(6)$ & $1.29(5)$ & $0.50(2)$ & $0.21$ \\
$2.00$ & $0.569(5)$ & $1.28(4)$ & $0.474(14)$ & $1.36$ \\
\hline
\hline
\end{tabular*}
\end{table}

\begin{figure}
\centerline{\epsfxsize=\columnwidth \epsfbox{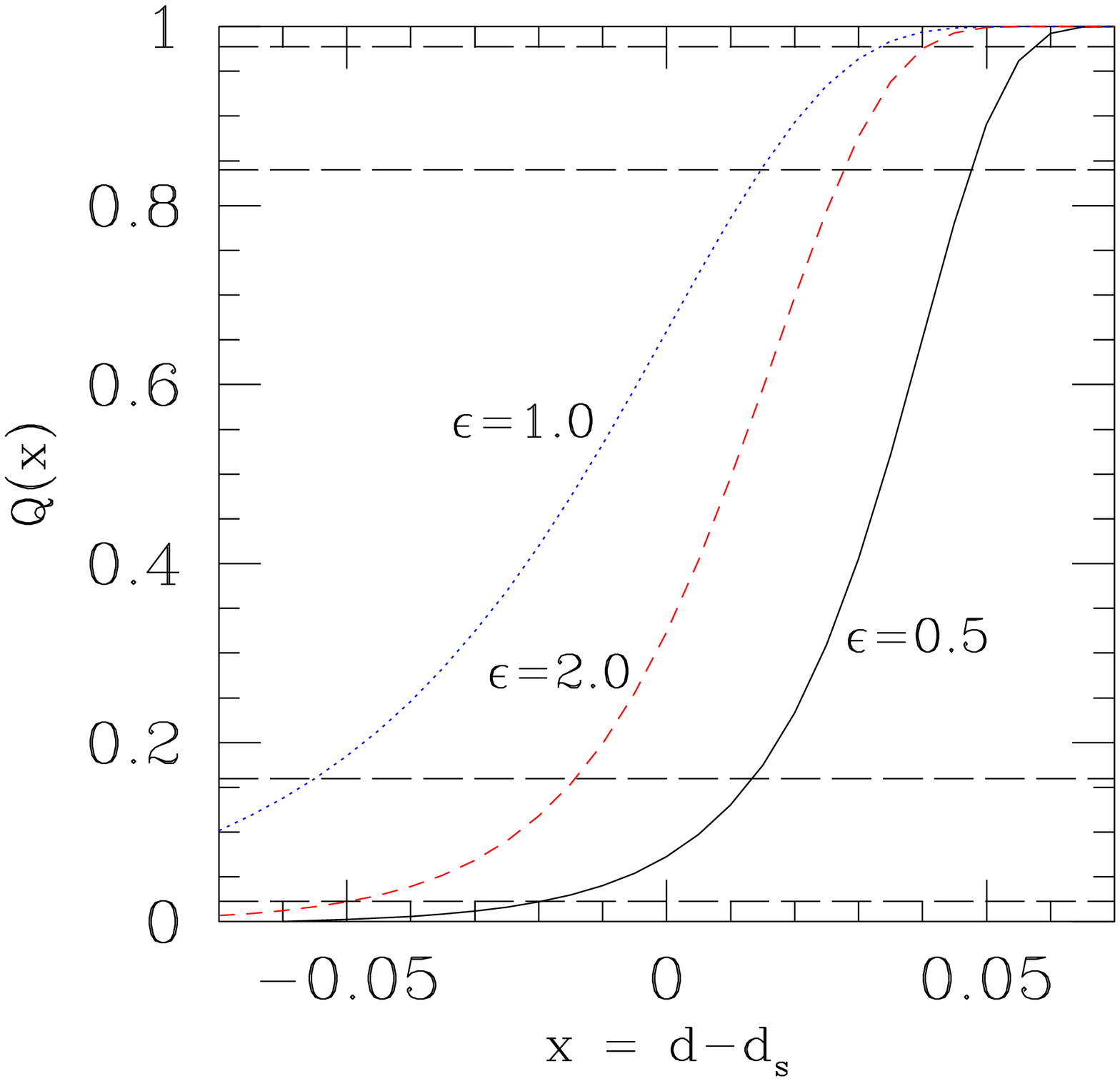}}
\vspace{-1.0cm}
\caption{
The cumulative probability for $d - d_s$ from the fits to $[1 - q_l]_{\rm
av}^\prime$ as discussed in the text. The inner pair of dashed horizontal
lines show $68\%$ confidence levels and the outer pair show $95.5\%$
confidence levels.
}
\label{cumul_1mqlp}
\end{figure}

Both DP/TNT and RSB fits are acceptable ($\chi^2/\mbox{ndf} \simeq 1$).
However, in fits to a nonlinear model, one cannot convert
$\chi^2/\mbox{ndf}$ to a confidence limit\cite{press:95} even if the data
have a normal distribution. Similarly, unlike for the case for fits to a
linear model, the error bars do not correspond to a $68\%$ confidence. We
are particularly interested to get a confidence limit on the value of $d -
d_s$ in the DP/TNT fits. We do this by computing $\chi^2$ as a function of
$d - d_s$, minimizing with respect to the other parameters ($a$, $b$, and
$c$). The probability of the fit $P$ is proportional to $\exp(-\chi^2/2)$
which we numerically integrate to get the cumulative probability for
$x = d-d_s$:
\begin{equation}
Q(x) = \int^x P(x^\prime) dx^\prime \; .
\label{Qx}
\end{equation}

The results are shown in Fig.~\ref{cumul_1mqlp}. The data for $\epsilon =
0.5$ and $2.0$ constrain $d-d_s$ to zero or a small positive value. The
data for $\epsilon = 1.0$ constrain $d - d_s$ less and allow a range of
negative values which are unphysical. At a $68\%$ confidence level the data
are consistent with
\begin{equation}
0 \le d - d_s \lesssim 0.05,
\label{dmds}
\end{equation}
apart from the $\epsilon = 0.5$ data which would exclude zero at the
$68\%$ level but, from Fig.~\ref{cumul_1mqlp}, are consistent with
it at the $86\%$ level. We take Eq.~(\ref{dmds})
to be our estimate for $d - d_s$. It is consistent with the RSB prediction
of zero and also consistent with a small non-zero value in the DP/TNT
scenarios.

\begin{figure}
\centerline{\epsfxsize=\columnwidth \epsfbox{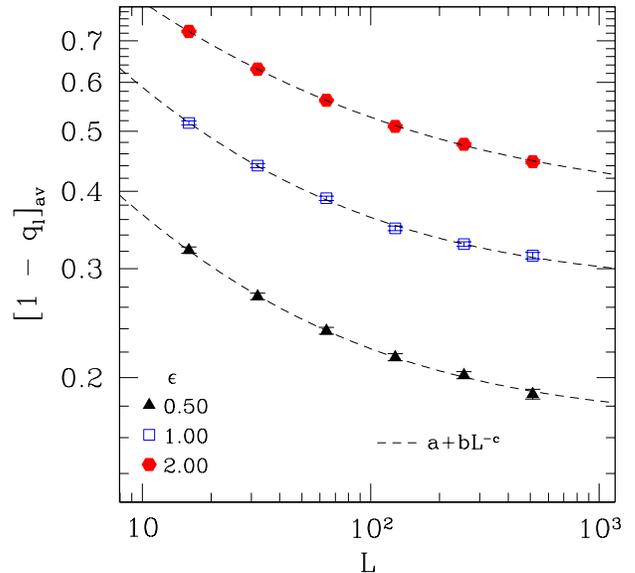}}
\vspace{-1.0cm}
\caption{
Zero-temperature data for $[1 - q_l]_{\rm av}$ as a function of system
size $L$ for different values of the coupling constant $\epsilon$. The
dashed lines represent fits according to $[1 - q_l]_{\rm av} = a +
bL^{-c}$ (RSB).
}
\label{omql_fig}
\end{figure}

Data for the average of $1 - q_l$ over \textit{all} samples are shown in
Fig.~\ref{omql_fig} along with RSB fits. The data show curvature
indicative that corrections to scaling have to be included. The DP/TNT
fits, Eq.~(\ref{ql_scale}), are shown in Table \ref{table_omqlfit1}, while
the RSB fits (which fix $\mu_l$ to zero) are shown in Table
\ref{table_omqlfit2}. Both fits have acceptable $\chi^2$.

\begin{figure}
\centerline{\epsfxsize=\columnwidth \epsfbox{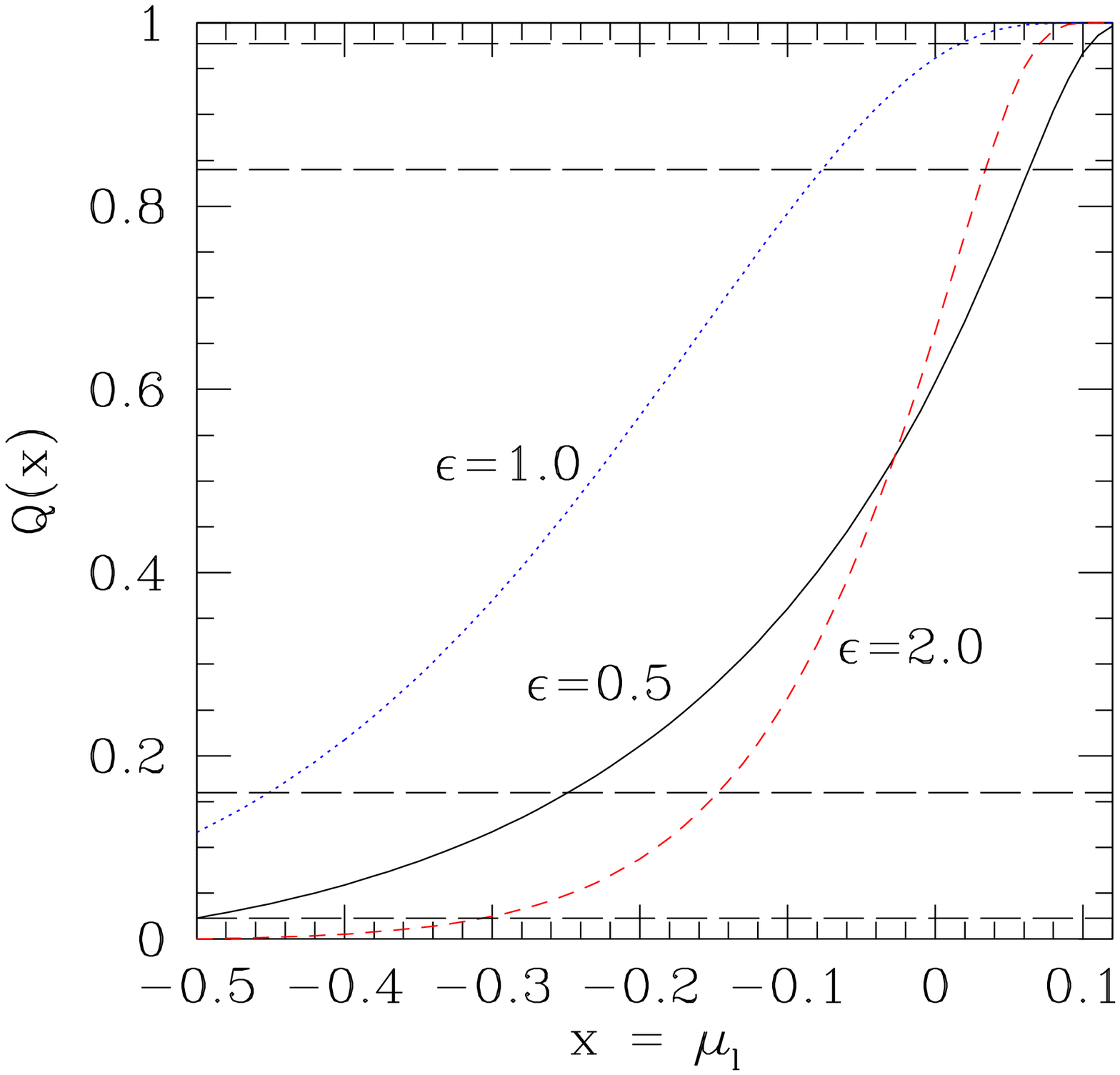}}
\vspace{-1.0cm}
\caption{
The cumulative probability for $d - d_s$ from the fits to
$[1 - q_l]_{\rm av}$ as
discussed in the text. The inner pair of dashed horizontal lines show $68\%$
confidence levels and the outer pair show $95.5\%$ confidence levels.
}
\label{cumul_1mql}
\end{figure}

The cumulative probabilities shown in Fig.~\ref{cumul_1mql} give a lot of
weight to unphysical negative values of $\mu_l$, especially for $\epsilon
= 1$. For all values of $\epsilon$ the weight is small for $\mu_l $
greater than about $0.10$ so we conclude that
\begin{equation}
0 \le \mu_l \lesssim 0.10 .
\label{mul}
\end{equation}
We should, perhaps, be cautious about this statement in view of the large
weight at negative values of $\mu_l$ in Fig.~\ref{cumul_1mql}. However,
Eq.~(\ref{mul}) is consistent with Eq.~(\ref{dmds}) and the result of
Ref.~\onlinecite{katzgraber:03} that $\theta^\prime \simeq 0$.

\begin{table}
\caption{
Fits of the zero-temperature data for $[1 - q_l]_{\rm av} = L^{-\mu_l}(a +
bL^{-c})$, which assume the DP/TNT picture, for different coupling
constants $\epsilon$.
\label{table_omqlfit1}
}
\begin{tabular*}{\columnwidth}{@{\extracolsep{\fill}} l l l l l l}
\hline
\hline
$\epsilon$ & $\mu_l$ & $a$ & $b$ & $c$ & $\chi^2 /\mbox{ndf}$\\
\hline
$0.50$ & $0.065(62)$ & $0.28(1)$  & $1.0(8)$  & $0.81(46)$ & $0.12$\\
$1.00$ &  \hspace{-2.55mm}$-0.15(18)$ & $0.07(12)$ & $1.07(7)$ & $0.51(6)$  & $0.68$\\
$2.00$ & $0.018(7)$  & $0.44(25)$ & $1.25(8)$ & $0.49(13)$ & $0.64$\\
\hline
\hline
\end{tabular*}
\end{table}

\begin{table}
\caption{
RSB fits of zero-temperature data for $[1 - q_l]_{\rm av} = a + bL^{-c}$
for different values of the coupling constant $\epsilon$.
\label{table_omqlfit2}
}
\begin{tabular*}{\columnwidth}{@{\extracolsep{\fill}} l l l l l}
\hline
\hline
$\epsilon$ & $a$ & $b$ & $c$ & $\chi^2 / \mbox{ndf}$\\
\hline
$0.50$ & $0.168(8)$ & $0.73(11)$  & $0.56(7)$ & $0.25$ \\
$1.00$ & $0.280(8)$ & $1.15(11)$ & $0.57(5)$ & $0.90$ \\
$2.00$ & $0.378(10)$ & $1.25(6)$  & $0.46(3)$ & $0.45$ \\
\hline
\hline
\end{tabular*}
\end{table}

\section{Results at Finite Temperature}
\label{Tgt0res}

\begin{figure}
\centerline{\epsfxsize=\columnwidth \epsfbox{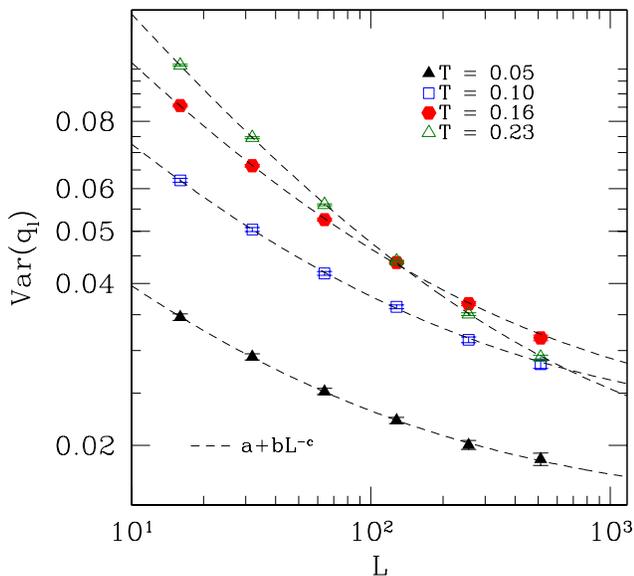}}
\vspace{-1.0cm}
\caption{
Log-log plot of finite-$T$ data for the variance of the link overlap ${\rm
Var}(q_l)$ as a function of system size $L$ for several low temperatures.
In all three cases we see strong curvature in the data suggesting
corrections to scaling. The dashed lines represent fits according to $a +
bL^{-c}$ (RSB) with the fitting parameters shown in Table
\ref{table_varqlfit2}. 
}
\label{varql_fig}
\end{figure}

\begin{figure}
\centerline{\epsfxsize=\columnwidth \epsfbox{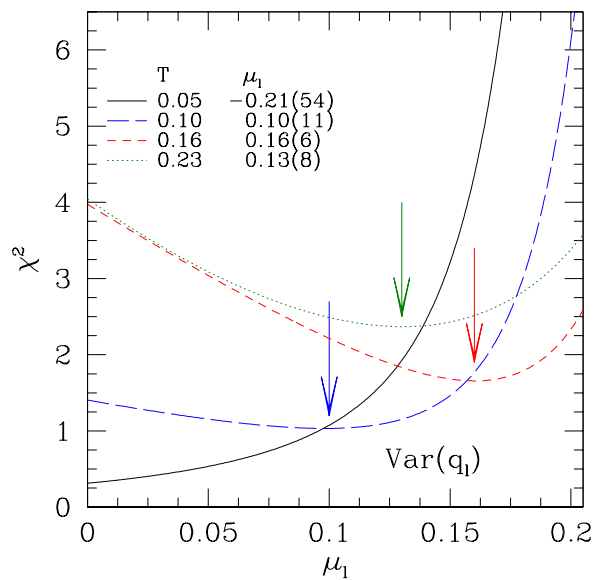}}
\vspace{-1.0cm}
\caption{
$\chi^2$ as a function of $\mu_l$, optimized with respect to the other
parameters ($a$, $b$, and $c$) in Eq.~(\ref{varql_scaling}), for the variance
of the link overlap. The arrows mark the minima.
}
\label{chi-sq_varql_fig}
\end{figure}

In this section we study the model at temperatures well
below\cite{katzgraber:03} $T_c \approx 0.63$. Figure \ref{varql_fig} shows
data for the variance of the link overlap for several low temperatures.
The data show strong curvature indicative that a simple fit of the form $a
L^{-\mu_l}$ is improbable and corrections to scaling must be included.

\begin{table}
\caption{
DP/TNT fits of ${\rm Var}(q_l)$ to $L^{-\mu_l}(a + bL^{-c})$ for different
temperatures.
\label{table_varqlfit1}
}
\begin{tabular*}{\columnwidth}{@{\extracolsep{\fill}} l l l l l l}
\hline
\hline
$T$ & $\mu_l$ & $a$ & $b$ & $c$ & $\chi^2 / \mbox{ndf}$\\
\hline
$0.05$ & \hspace{-2.55mm}$-0.21(54)$ & $0.002(11)$ & $0.073(8)$ & $0.52(35)$ & $0.05$\\
$0.10$ & $ 0.10(11)$ & $0.047(42)$ & $0.16(3)$  & $0.55(24)$ & $0.52$\\
$0.16$ & $ 0.16(6)$  & $0.079(39)$ & $0.29(6)$  & $0.60(17)$ & $0.83$\\
$0.23$ & $ 0.13(8)$  & $0.050(34)$ & $0.41(2)$  & $0.52(2)$  & $1.18$\\
\hline
\hline
\end{tabular*}
\end{table}

\begin{table}
\caption{
RSB fits of ${\rm Var}(q_l)$ to $ a + bL^{-c}$ for different temperatures.
\label{table_varqlfit2}
}
\begin{tabular*}{\columnwidth}{@{\extracolsep{\fill}} l l l l l}
\hline
\hline
$T$ & $a$ & $b$ & $c$ & $\chi^2 / \mbox{ndf}$\\
\hline
$0.05$ & $0.015(2)$ & $0.073(10)$ & $0.47(7)$ & $0.10$ \\
$0.10$ & $0.020(2)$ & $0.155(11)$ & $0.47(3)$ & $0.47$ \\
$0.16$ & $0.021(1)$ & $0.261(11)$ & $0.50(2)$ & $1.32$ \\
$0.23$ & $0.017(1)$ & $0.391(9)$  & $0.55(1)$ & $1.35$ \\
\hline
\hline
\end{tabular*}
\end{table}

Fits to Eq.~(\ref{varql_scaling}) (DP/TNT picture) are shown in Table
\ref{table_varqlfit1}, and fits to the RSB picture (in which $\mu_l$ is
fixed to be zero) are shown in Table \ref{table_varqlfit2}. The quality of
the fits is acceptable.

However, the DP/TNT fit for $T = 0.05$ gives an unphysical negative value
for $\mu_l$ with a very small amplitude $a$. To clarify this situation, we
plot, in Fig.~\ref{chi-sq_varql_fig}, $\chi^2$ as a function of $\mu_l$,
optimizing with respect to the other parameters ($a, b$, and $c$).  For
$T=0.05$, $\chi^2$ is quite small out to very large \textit{negative}
values of $\mu_l$ (not shown) and increases rapidly for $\mu_l$ greater
than about $0.12$. Since physically $\mu_l$ cannot be negative, the only
conclusion we can deduce from the $T=0.05$ data is that $\mu_l$ lies
between zero and about $0.12$, consistent with the result from the $T=0$
data in Eq.~(\ref{mul}). The data for $\chi^2$ for $T=0.10$ in
Fig.~\ref{chi-sq_varql_fig} has a minimum at $\mu_l = 0.10$ but it is
shallow and $\mu_l = 0$ has only a slightly greater $\chi^2$ value. The 
$T=0.10$ data are therefore also consistent with Eq.~(\ref{mul}). The data at
higher temperatures, $T = 0.16$ and $0.23$ have a deeper minimum at
nonzero $\chi^2$, suggesting that $\mu_l = 0$ is somewhat unlikely, but
experience from short-range systems\cite{katzgraber:01} suggests that
estimates of $\mu_l$ at finite $T$ are effective exponents which need to
be extrapolated to $T = 0$ to get close to the asymptotic value. Hence we
do not feel that the results at $T = 0.16$ and $0.23$ rule out $\mu_l =
0$.

Overall, the finite-$T$ data are consistent with $\mu_l$ in the range given
by Eq.~(\ref{mul}) which came from the $T=0$ data, and do not constrain
$\mu_l$ any further.

\section{Conclusions}
\label{conclusions}

We have studied the geometry of the large-scale, low-energy excitations in
a one-dimensional Ising spin glass where the interactions fall off as
$r^{-\sigma}$ with $\sigma = 0.75$, both at $T=0$ and at temperatures well
below the spin-glass transition temperature. We find that the fractal
dimension of the surface of these excitations, $d_s$, lies in the range
$0 \le d - d_s \lesssim 0.05$. This is consistent with the RSB picture ($d -
d_s = 0$). It is also consistent with the DP/TNT picture ($d - d_s > 0$)
but with a small value of $d-d_s$. Substantial corrections to scaling had
to be incorporated into all the fits.

We have also estimated the exponent $\mu_l = \theta^\prime + 2(d - d_s)$,
where $\theta^\prime$ characterizes the dependence of the energy of
droplet excitations on their length scale. We find it to be in the range
$0 \le \mu_l \lesssim 0.10$, which is consistent with the value for $d-d_s$
in Eq.~(\ref{dmds}) and our earlier result\cite{katzgraber:03} that
$\theta^\prime \simeq 0$. Note that this result for $\theta^\prime$, if also
valid in the thermodynamic limit, is inconsistent with the DP.

By studying a one-dimensional model, we have been able to study a much
larger range of sizes, $16 \le L \le 512$, than is generally possible in
spin glasses. However, in the absence of a good understanding of
corrections to scaling in spin glasses, we still cannot rule out the
possibility that different behavior \textit{may} occur in the
thermodynamic limit.

\begin{acknowledgments}
We would like to thank K.~Tran for carefully reading the manuscript. The
simulations were performed on the Asgard cluster at ETH Z\"urich. We are
indebted to M.~Troyer and G.~Sigut for allowing us to use the idle time on
the Asgard cluster. The work of APY is supported by the National Science
Foundation under Grant No.~DMR 0086287. \end{acknowledgments}

\bibliography{refs}

\begin{thebibliography}{33}
\expandafter\ifx\csname natexlab\endcsname\relax\def\natexlab#1{#1}\fi
\expandafter\ifx\csname bibnamefont\endcsname\relax
  \def\bibnamefont#1{#1}\fi
\expandafter\ifx\csname bibfnamefont\endcsname\relax
  \def\bibfnamefont#1{#1}\fi
\expandafter\ifx\csname citenamefont\endcsname\relax
  \def\citenamefont#1{#1}\fi
\expandafter\ifx\csname url\endcsname\relax
  \def\url#1{\texttt{#1}}\fi
\expandafter\ifx\csname urlprefix\endcsname\relax\def\urlprefix{URL }\fi
\providecommand{\bibinfo}[2]{#2}
\providecommand{\eprint}[2][]{\url{#2}}

\bibitem[{\citenamefont{Krzakala and Martin}(2000)}]{krzakala:00}
\bibinfo{author}{\bibfnamefont{F.}~\bibnamefont{Krzakala}} \bibnamefont{and}
  \bibinfo{author}{\bibfnamefont{O.~C.} \bibnamefont{Martin}},
  \emph{\bibinfo{title}{Spin and link overlaps in 3-dimensional spin glasses}},
  \bibinfo{journal}{Phys. Rev. Lett.} \textbf{\bibinfo{volume}{85}},
  \bibinfo{pages}{3013} (\bibinfo{year}{2000}),
  \bibinfo{note}{(cond-mat/0002055)}.

\bibitem[{\citenamefont{Palassini and Young}(2000)}]{palassini:00}
\bibinfo{author}{\bibfnamefont{M.}~\bibnamefont{Palassini}} \bibnamefont{and}
  \bibinfo{author}{\bibfnamefont{A.~P.} \bibnamefont{Young}},
  \emph{\bibinfo{title}{Nature of the spin glass state}},
  \bibinfo{journal}{Phys. Rev. Lett.} \textbf{\bibinfo{volume}{85}},
  \bibinfo{pages}{3017} (\bibinfo{year}{2000}),
  \bibinfo{note}{(cond-mat/0002134)}.

\bibitem[{\citenamefont{Marinari and Parisi}(2000)}]{marinari:00}
\bibinfo{author}{\bibfnamefont{E.}~\bibnamefont{Marinari}} \bibnamefont{and}
  \bibinfo{author}{\bibfnamefont{G.}~\bibnamefont{Parisi}},
  \emph{\bibinfo{title}{On the effects of changing the boundary conditions on
  the ground state of {I}sing spin glasses}}, \bibinfo{journal}{Phys. Rev. B}
  \textbf{\bibinfo{volume}{62}}, \bibinfo{pages}{11677} (\bibinfo{year}{2000}).

\bibitem[{\citenamefont{Katzgraber et~al.}(2001)\citenamefont{Katzgraber,
  Palassini, and Young}}]{katzgraber:01}
\bibinfo{author}{\bibfnamefont{H.~G.} \bibnamefont{Katzgraber}},
  \bibinfo{author}{\bibfnamefont{M.}~\bibnamefont{Palassini}},
  \bibnamefont{and} \bibinfo{author}{\bibfnamefont{A.~P.} \bibnamefont{Young}},
  \emph{\bibinfo{title}{{M}onte {C}arlo simulations of spin glasses at low
  temperatures}}, \bibinfo{journal}{Phys. Rev. B}
  \textbf{\bibinfo{volume}{63}}, \bibinfo{pages}{184422}
  (\bibinfo{year}{2001}), \bibinfo{note}{(cond-mat/0007113)}.

\bibitem[{\citenamefont{Katzgraber and Young}(2001)}]{katzgraber:01a}
\bibinfo{author}{\bibfnamefont{H.~G.} \bibnamefont{Katzgraber}}
  \bibnamefont{and} \bibinfo{author}{\bibfnamefont{A.~P.} \bibnamefont{Young}},
  \emph{\bibinfo{title}{Nature of the spin-glass state in the three-dimensional
  gauge glass}}, \bibinfo{journal}{Phys. Rev. B} \textbf{\bibinfo{volume}{64}},
  \bibinfo{pages}{104426} (\bibinfo{year}{2001}),
  \bibinfo{note}{(cond-mat/0105077)}.

\bibitem[{\citenamefont{Katzgraber and Young}(2002)}]{katzgraber:01c}
\bibinfo{author}{\bibfnamefont{H.~G.} \bibnamefont{Katzgraber}}
  \bibnamefont{and} \bibinfo{author}{\bibfnamefont{A.~P.} \bibnamefont{Young}},
  \emph{\bibinfo{title}{{M}onte {C}arlo simulations of the four dimensional
  {XY} spin-glass at low temperatures}}, \bibinfo{journal}{Phys. Rev. B}
  \textbf{\bibinfo{volume}{65}}, \bibinfo{pages}{214401}
  (\bibinfo{year}{2002}), \bibinfo{note}{(cond-mat/0108320)}.

\bibitem[{\citenamefont{Houdayer et~al.}(2000)\citenamefont{Houdayer, Krzakala,
  and Martin}}]{houdayer:00}
\bibinfo{author}{\bibfnamefont{J.}~\bibnamefont{Houdayer}},
  \bibinfo{author}{\bibfnamefont{F.}~\bibnamefont{Krzakala}}, \bibnamefont{and}
  \bibinfo{author}{\bibfnamefont{O.~C.} \bibnamefont{Martin}},
  \emph{\bibinfo{title}{Large-scale low-energy excitations in 3-d spin
  glasses}}, \bibinfo{journal}{Eur. Phys. J. B.} \textbf{\bibinfo{volume}{18}},
  \bibinfo{pages}{467} (\bibinfo{year}{2000}).

\bibitem[{\citenamefont{Houdayer and Martin}(2000)}]{houdayer:00b}
\bibinfo{author}{\bibfnamefont{J.}~\bibnamefont{Houdayer}} \bibnamefont{and}
  \bibinfo{author}{\bibfnamefont{O.~C.} \bibnamefont{Martin}},
  \emph{\bibinfo{title}{{A} geometric picture for finite dimensional spin
  glasses}}, \bibinfo{journal}{Europhys. Lett.} \textbf{\bibinfo{volume}{49}},
  \bibinfo{pages}{794} (\bibinfo{year}{2000}).

\bibitem[{\citenamefont{Bray and Moore}(1984)}]{bray:84}
\bibinfo{author}{\bibfnamefont{A.~J.} \bibnamefont{Bray}} \bibnamefont{and}
  \bibinfo{author}{\bibfnamefont{M.~A.} \bibnamefont{Moore}},
  \emph{\bibinfo{title}{Lower critical dimension of {I}sing spin glasses: a
  numerical study}}, \bibinfo{journal}{J. Phys. C}
  \textbf{\bibinfo{volume}{17}}, \bibinfo{pages}{L463} (\bibinfo{year}{1984}).

\bibitem[{\citenamefont{McMillan}(1984{\natexlab{a}})}]{mcmillan:84}
\bibinfo{author}{\bibfnamefont{W.~L.} \bibnamefont{McMillan}},
  \emph{\bibinfo{title}{Domain-wall renormalization-group study of the
  three-dimensional random {I}sing model}}, \bibinfo{journal}{Phys. Rev. B}
  \textbf{\bibinfo{volume}{30}}, \bibinfo{pages}{476}
  (\bibinfo{year}{1984}{\natexlab{a}}).

\bibitem[{\citenamefont{McMillan}(1984{\natexlab{b}})}]{mcmillan:84b}
\bibinfo{author}{\bibfnamefont{W.~L.} \bibnamefont{McMillan}},
  \emph{\bibinfo{title}{Domain-wall renormalization-group study of the
  two-dimensional random {I}sing model}}, \bibinfo{journal}{Phys. Rev. B}
  \textbf{\bibinfo{volume}{29}}, \bibinfo{pages}{4026}
  (\bibinfo{year}{1984}{\natexlab{b}}).

\bibitem[{\citenamefont{Rieger et~al.}(1996)\citenamefont{Rieger, Santen,
  Blasum, Diehl, J\"unger, and Rinaldi}}]{rieger:96}
\bibinfo{author}{\bibfnamefont{H.}~\bibnamefont{Rieger}},
  \bibinfo{author}{\bibfnamefont{L.}~\bibnamefont{Santen}},
  \bibinfo{author}{\bibfnamefont{U.}~\bibnamefont{Blasum}},
  \bibinfo{author}{\bibfnamefont{M.}~\bibnamefont{Diehl}},
  \bibinfo{author}{\bibfnamefont{M.}~\bibnamefont{J\"unger}}, \bibnamefont{and}
  \bibinfo{author}{\bibfnamefont{G.}~\bibnamefont{Rinaldi}},
  \emph{\bibinfo{title}{The critical exponents of the two-dimensional {I}sing
  spin glass revisited: exact ground-state calculations and {M}onte {C}arlo
  simulations}}, \bibinfo{journal}{J. Phys. A} \textbf{\bibinfo{volume}{29}},
  \bibinfo{pages}{3939} (\bibinfo{year}{1996}).

\bibitem[{\citenamefont{Hartmann}(1999)}]{hartmann:99}
\bibinfo{author}{\bibfnamefont{A.~K.} \bibnamefont{Hartmann}},
  \emph{\bibinfo{title}{Scaling of stiffness energy for three-dimensional $\pm
  {J}$ {I}sing spin glasses}}, \bibinfo{journal}{Phys. Rev. E}
  \textbf{\bibinfo{volume}{59}}, \bibinfo{pages}{84} (\bibinfo{year}{1999}).

\bibitem[{\citenamefont{Palassini and Young}(1999)}]{palassini:99}
\bibinfo{author}{\bibfnamefont{M.}~\bibnamefont{Palassini}} \bibnamefont{and}
  \bibinfo{author}{\bibfnamefont{A.~P.} \bibnamefont{Young}},
  \emph{\bibinfo{title}{Triviality of the ground state structure in {I}sing
  spin glasses}}, \bibinfo{journal}{Phys. Rev. Lett.}
  \textbf{\bibinfo{volume}{83}}, \bibinfo{pages}{5126} (\bibinfo{year}{1999}),
  \bibinfo{note}{(cond-mat/9906323)}.

\bibitem[{\citenamefont{Hartmann and Young}(2001)}]{hartmann:01a}
\bibinfo{author}{\bibfnamefont{A.~K.} \bibnamefont{Hartmann}} \bibnamefont{and}
  \bibinfo{author}{\bibfnamefont{A.~P.} \bibnamefont{Young}},
  \emph{\bibinfo{title}{Lower critical dimension of {I}sing spin glasses}},
  \bibinfo{journal}{Phys. Rev. B} \textbf{\bibinfo{volume}{64}},
  \bibinfo{pages}{180404} (\bibinfo{year}{2001}),
  \bibinfo{note}{(cond-mat/0107308)}.

\bibitem[{\citenamefont{Carter et~al.}(2002)\citenamefont{Carter, Bray, and
  Moore}}]{carter:02}
\bibinfo{author}{\bibfnamefont{A.~C.} \bibnamefont{Carter}},
  \bibinfo{author}{\bibfnamefont{A.~J.} \bibnamefont{Bray}}, \bibnamefont{and}
  \bibinfo{author}{\bibfnamefont{M.~A.} \bibnamefont{Moore}},
  \emph{\bibinfo{title}{Aspect-ratio scaling and the stiffness exponent
  $\theta$ for {I}sing spin glasses}}, \bibinfo{journal}{Phys. Rev. Lett.}
  \textbf{\bibinfo{volume}{88}}, \bibinfo{pages}{077201}
  (\bibinfo{year}{2002}), \bibinfo{note}{(cond-mat/0108050)}.

\bibitem[{\citenamefont{Parisi}(1979)}]{parisi:79}
\bibinfo{author}{\bibfnamefont{G.}~\bibnamefont{Parisi}},
  \emph{\bibinfo{title}{Infinite number of order parameters for spin-glasses}},
  \bibinfo{journal}{Phys. Rev. Lett.} \textbf{\bibinfo{volume}{43}},
  \bibinfo{pages}{1754} (\bibinfo{year}{1979}).

\bibitem[{\citenamefont{Parisi}(1980)}]{parisi:80}
\bibinfo{author}{\bibfnamefont{G.}~\bibnamefont{Parisi}},
  \emph{\bibinfo{title}{The order parameter for spin glasses: a function on the
  interval $0$--$1$}}, \bibinfo{journal}{J. Phys. A}
  \textbf{\bibinfo{volume}{13}}, \bibinfo{pages}{1101} (\bibinfo{year}{1980}).

\bibitem[{\citenamefont{Parisi}(1983)}]{parisi:83}
\bibinfo{author}{\bibfnamefont{G.}~\bibnamefont{Parisi}},
  \emph{\bibinfo{title}{Order parameter for spin-glasses}},
  \bibinfo{journal}{Phys. Rev. Lett.} \textbf{\bibinfo{volume}{50}},
  \bibinfo{pages}{1946} (\bibinfo{year}{1983}).

\bibitem[{\citenamefont{M\'ezard et~al.}(1987)\citenamefont{M\'ezard, Parisi,
  and Virasoro}}]{mezard:87}
\bibinfo{author}{\bibfnamefont{M.}~\bibnamefont{M\'ezard}},
  \bibinfo{author}{\bibfnamefont{G.}~\bibnamefont{Parisi}}, \bibnamefont{and}
  \bibinfo{author}{\bibfnamefont{M.~A.} \bibnamefont{Virasoro}},
  \emph{\bibinfo{title}{Spin Glass Theory and Beyond}}
  (\bibinfo{publisher}{World Scientific}, \bibinfo{address}{Singapore},
  \bibinfo{year}{1987}).

\bibitem[{\citenamefont{Fisher and Huse}(1986)}]{fisher:86}
\bibinfo{author}{\bibfnamefont{D.~S.} \bibnamefont{Fisher}} \bibnamefont{and}
  \bibinfo{author}{\bibfnamefont{D.~A.} \bibnamefont{Huse}},
  \emph{\bibinfo{title}{Ordered phase of short-range {I}sing spin-glasses}},
  \bibinfo{journal}{Phys. Rev. Lett.} \textbf{\bibinfo{volume}{56}},
  \bibinfo{pages}{1601} (\bibinfo{year}{1986}).

\bibitem[{\citenamefont{Fisher and Huse}(1987)}]{fisher:87}
\bibinfo{author}{\bibfnamefont{D.~S.} \bibnamefont{Fisher}} \bibnamefont{and}
  \bibinfo{author}{\bibfnamefont{D.~A.} \bibnamefont{Huse}},
  \emph{\bibinfo{title}{Absence of many states in realistic spin glasses}},
  \bibinfo{journal}{J. Phys. A} \textbf{\bibinfo{volume}{20}},
  \bibinfo{pages}{L1005} (\bibinfo{year}{1987}).

\bibitem[{\citenamefont{Fisher and Huse}(1988)}]{fisher:88}
\bibinfo{author}{\bibfnamefont{D.~S.} \bibnamefont{Fisher}} \bibnamefont{and}
  \bibinfo{author}{\bibfnamefont{D.~A.} \bibnamefont{Huse}},
  \emph{\bibinfo{title}{Equilibrium behavior of the spin-glass ordered phase}},
  \bibinfo{journal}{Phys. Rev. B} \textbf{\bibinfo{volume}{38}},
  \bibinfo{pages}{386} (\bibinfo{year}{1988}).

\bibitem[{\citenamefont{Bray and Moore}(1986)}]{bray:86}
\bibinfo{author}{\bibfnamefont{A.~J.} \bibnamefont{Bray}} \bibnamefont{and}
  \bibinfo{author}{\bibfnamefont{M.~A.} \bibnamefont{Moore}},
  \emph{\bibinfo{title}{Scaling theory of the ordered phase of spin glasses}},
  in \emph{\bibinfo{booktitle}{Heidelberg Colloquium on Glassy Dynamics and
  Optimization}}, edited by
  \bibinfo{editor}{\bibfnamefont{L.}~\bibnamefont{Van~Hemmen}}
  \bibnamefont{and}
  \bibinfo{editor}{\bibfnamefont{I.}~\bibnamefont{Morgenstern}}
  (\bibinfo{publisher}{Springer}, \bibinfo{address}{New York},
  \bibinfo{year}{1986}), p. \bibinfo{pages}{121}.

\bibitem[{\citenamefont{Katzgraber and Young}(2003)}]{katzgraber:03}
\bibinfo{author}{\bibfnamefont{H.~G.} \bibnamefont{Katzgraber}}
  \bibnamefont{and} \bibinfo{author}{\bibfnamefont{A.~P.} \bibnamefont{Young}},
  \emph{\bibinfo{title}{Monte {C}arlo studies of the one-dimensional {I}sing
  spin glass with power-law interactions}}, \bibinfo{journal}{Phys. Rev. B}
  \textbf{\bibinfo{volume}{67}}, \bibinfo{pages}{134410}
  (\bibinfo{year}{2003}), \bibinfo{note}{(cond-mat/0210451)}.

\bibitem[{\citenamefont{Bray et~al.}(1986)\citenamefont{Bray, Moore, and
  Young}}]{bray:86b}
\bibinfo{author}{\bibfnamefont{A.~J.} \bibnamefont{Bray}},
  \bibinfo{author}{\bibfnamefont{M.~A.} \bibnamefont{Moore}}, \bibnamefont{and}
  \bibinfo{author}{\bibfnamefont{A.~P.} \bibnamefont{Young}},
  \emph{\bibinfo{title}{Lower critical dimension of metallic vector
  spin-glasses}}, \bibinfo{journal}{Phys. Rev. Lett}
  \textbf{\bibinfo{volume}{56}}, \bibinfo{pages}{2641} (\bibinfo{year}{1986}).

\bibitem[{\citenamefont{Palassini et~al.}(2002)\citenamefont{Palassini, Liers,
  J\"unger, and Young}}]{palassini:02}
\bibinfo{author}{\bibfnamefont{M.}~\bibnamefont{Palassini}},
  \bibinfo{author}{\bibfnamefont{F.}~\bibnamefont{Liers}},
  \bibinfo{author}{\bibfnamefont{M.}~\bibnamefont{J\"unger}}, \bibnamefont{and}
  \bibinfo{author}{\bibfnamefont{A.~P.} \bibnamefont{Young}},
  \emph{\bibinfo{title}{Structure of the spin glass phase from perturbed exact
  ground states}} (\bibinfo{year}{2002}), \bibinfo{note}{(cond-mat/0212551)}.

\bibitem[{\citenamefont{Hartmann and Young}(2002)}]{hartmann:02}
\bibinfo{author}{\bibfnamefont{A.~K.} \bibnamefont{Hartmann}} \bibnamefont{and}
  \bibinfo{author}{\bibfnamefont{A.~P.} \bibnamefont{Young}},
  \emph{\bibinfo{title}{Large-scale low-energy excitations in the
  two-dimensional {I}sing spin glass}}, \bibinfo{journal}{Phys. Rev. B}
  \textbf{\bibinfo{volume}{66}}, \bibinfo{pages}{094419}
  (\bibinfo{year}{2002}), \bibinfo{note}{(cond-mat/0205659)}.

\bibitem[{\citenamefont{Hukushima and Nemoto}(1996)}]{hukushima:96}
\bibinfo{author}{\bibfnamefont{K.}~\bibnamefont{Hukushima}} \bibnamefont{and}
  \bibinfo{author}{\bibfnamefont{K.}~\bibnamefont{Nemoto}},
  \emph{\bibinfo{title}{Exchange {M}onte {C}arlo method and application to spin
  glass simulations}}, \bibinfo{journal}{J. Phys. Soc. Jpn.}
  \textbf{\bibinfo{volume}{65}}, \bibinfo{pages}{1604} (\bibinfo{year}{1996}).

\bibitem[{\citenamefont{Marinari}(1998)}]{marinari:98b}
\bibinfo{author}{\bibfnamefont{E.}~\bibnamefont{Marinari}},
  \emph{\bibinfo{title}{Optimized {M}onte {C}arlo methods}}, in
  \emph{\bibinfo{booktitle}{Advances in Computer Simulation}}, edited by
  \bibinfo{editor}{\bibfnamefont{J.}~\bibnamefont{Kert\'esz}} \bibnamefont{and}
  \bibinfo{editor}{\bibfnamefont{I.}~\bibnamefont{Kondor}}
  (\bibinfo{publisher}{Springer-Verlag}, \bibinfo{address}{Berlin},
  \bibinfo{year}{1998}), p.~\bibinfo{pages}{50},
  \bibinfo{note}{(cond-mat/9612010)}.

\bibitem[{\citenamefont{Moreno et~al.}(2003)\citenamefont{Moreno, Katzgraber,
  and Hartmann}}]{moreno:02}
\bibinfo{author}{\bibfnamefont{J.~J.} \bibnamefont{Moreno}},
  \bibinfo{author}{\bibfnamefont{H.~G.} \bibnamefont{Katzgraber}},
  \bibnamefont{and} \bibinfo{author}{\bibfnamefont{A.~K.}
  \bibnamefont{Hartmann}}, \emph{\bibinfo{title}{Finding low-temperature states
  with parallel tempering, simulated annealing and simple {M}onte {C}arlo}},
  \bibinfo{journal}{Int. J. Mod. Phys. C} \textbf{\bibinfo{volume}{14}},
  \bibinfo{pages}{285} (\bibinfo{year}{2003}),
  \bibinfo{note}{(cond-mat/0209248)}.

\bibitem[{\citenamefont{Leuzzi}(1999)}]{leuzzi:99}
\bibinfo{author}{\bibfnamefont{L.}~\bibnamefont{Leuzzi}},
  \emph{\bibinfo{title}{Critical behaviour and ultrametricity of ising
  spin-glass with long-range interactions}}, \bibinfo{journal}{J. Phys. A}
  \textbf{\bibinfo{volume}{32}}, \bibinfo{pages}{1417} (\bibinfo{year}{1999}).

\bibitem[{\citenamefont{Press et~al.}(1995)\citenamefont{Press, Teukolsky,
  Vetterling, and Flannery}}]{press:95}
\bibinfo{author}{\bibfnamefont{W.~H.} \bibnamefont{Press}},
  \bibinfo{author}{\bibfnamefont{S.~A.} \bibnamefont{Teukolsky}},
  \bibinfo{author}{\bibfnamefont{W.~T.} \bibnamefont{Vetterling}},
  \bibnamefont{and} \bibinfo{author}{\bibfnamefont{B.~P.}
  \bibnamefont{Flannery}}, \emph{\bibinfo{title}{Numerical Recipes in C}}
  (\bibinfo{publisher}{Cambridge University Press},
  \bibinfo{address}{Cambridge}, \bibinfo{year}{1995}).

\end{thebibliography}

\end{document}